\documentclass[twocolumn,english,aps,tightenlines,showpacs,showkeys,notitlepage]{revtex4-2}
\usepackage[latin9]{inputenc}
\usepackage{babel}
\usepackage{amssymb}
\usepackage{esint}
\usepackage[unicode=true,pdfusetitle,
 bookmarks=true,bookmarksnumbered=false,bookmarksopen=false,
 breaklinks=false,pdfborder={0 0 1},backref=false,colorlinks=false]
 {hyperref}

\makeatletter

\DeclareFontEncoding{LGR}{}{}
\DeclareRobustCommand{\greektext}{%
  \fontencoding{LGR}\selectfont\def\encodingdefault{LGR}}
\DeclareRobustCommand{\textgreek}[1]{\leavevmode{\greektext #1}}
\ProvideTextCommand{\~}{LGR}[1]{\char126#1}

\usepackage{dcolumn}
\usepackage{bm}

\usepackage{babel}

\makeatother

\begin{document}
\title{GUP-corrected \textgreek{L}CDM cosmology}
\author{Salih Kibaro\u{g}lu}
\email{salihkibaroglu@maltepe.edu.tr}

\date{\today}
\begin{abstract}
In this study, we investigate the effect of the generalized uncertainty
principle on the \textgreek{L}CDM cosmological model. Using quantum
corrected Unruh effect and Verlinde\textquoteright s entropic gravity
idea, we find Planck-scale corrected Friedmann equations with a cosmological
constant. These results modify the \textgreek{L}CDM cosmology.
\end{abstract}
\affiliation{T.C. Maltepe University, Faculty of Engineering and Natural Sciences,
Istanbul, Turkey }
\keywords{Generalized gravity theories, cosmology, uncertainty principle}
\maketitle

\section{Introduction}

In general, the dynamics of our Universe is explained by cosmological
models. But there is no cosmological model that explains all phases
of the evolution of the Universe. According to the Friedmann-Lemaitre-Robertson-Walker
model, the Universe is assumed that it has a homogeneous and isotropic
characteristic on a large scale. Furthermore, we know that this is
an approximate theory and we should find more general cosmological
models to explain our Universe more accurate. 

In recent years, the \textgreek{L}-cold dark matter (\textgreek{L}CDM)
cosmological model is known as the standard model of cosmology. According
to this model, Einstein\textquoteright s theory of general relativity
is a correct theory and the Universe has a characteristic like an
Einstein-de Sitter space-time for the late time. This model provides
well explanations from today to the radiation-dominated era such as
the accelerated expansion of the Universe, cosmic microwave background.
On the other hand, this theory is not successful to explain the early
time period of the Universe such as the inflation era. For this reason,
it is reasonable to extend/modify this theory to a more general form. 

At this point, the extended theories of gravity help us to find generalized
cosmological models. For instance, in 1998 it was discovered that
the expansion rate of our Universe is accelerating \citep{perlmutter1998discovery,perlmutter1999measurements,riess1998observational,riess2004type}.
To explain this acceleration, the main candidates are the cosmological
constant and dark energy which can be obtained by adding extra source
terms to the Einstein field equations \citep{frieman2008dark,padmanabhan2009dark}. 

There is another alternative approach for constructing an extended
theory of gravity put forwarded by Verlinde \citep{Verlinde:2010hp,Verlinde:2016toy}.
According to his theory, gravity is not a fundamental force but emerges
as an entropic force. He formulated a thermodynamical description
of gravity by considering the holographic principle and Bekenstein-Hawking
entropy \citep{Bekenstein1973,Hawking:1974rv,Hawking:1975vcx}. In
the light of this idea, if one has modified/deformed thermodynamical
quantity this leads to finding modified gravitational equations. This
idea provides a very useful theoretical background to derive generalized
versions of gravity theories \citep{Ho:2010ca,Wang:2012gc,Sheykhi:2010yq,Sheykhi:2012vf,Dil:2015mgl,Moradpour:2015pem,Senay:2018xaj,Kibaroglu:2018mnx,Kibaroglu:2019dwd}
and cosmological models \citep{Cai:2010hk,Cai:2010zw,Sheykhi:2010wm,Wei:2010am,Cai:2010kp,Komatsu:2012zh,Komatsu:2013qia,Awad:2014nma,Sheykhi:2018dpn,Kibaroglu:2019odt}. 

The quantum gravity studies show that Heisenberg\textquoteright s
uncertainty principle should be modified \citep{Kempf:1994su}. In
this context, one of the main studies is the Generalized Uncertainty
Principle (GUP) \citep{Garay:1994en,Scardigli:1995qd,Scardigli:1999jh,KalyanaRama:2001xd,Chang:2001bm,Hossenfelder:2012jw,Gine:2018tjn,Scardigli:2018jlm}
which provide a quantum gravitational correction to the Heisenberg
relation as, 

\begin{equation}
\Delta x\Delta p\geq\frac{\hbar}{2}\left[1+\beta l_{p}^{2}\left(\Delta p\right)^{2}\right],
\end{equation}
where $l_{p}$ is the Planck length, $\beta$ is a constant of order
unity and dimensionless and $c=1$ is assumed. Furthermore, another
study is the Extended Uncertainty Principle (EUP) have a position-uncertainty
correction to the ordinary uncertainty relation, 

\begin{equation}
\Delta x\Delta p\geq i\frac{\hbar}{2}\left[1+\frac{\alpha}{l_{H}^{2}}\left(\Delta x\right)^{2}\right],
\end{equation}
where $\alpha$ is taken to be of order unity and $l_{H}$ is a large
distance scale for instance the (anti)-de Sitter radius. Thus, EUP
provides a way to introduce quantum effects on large scales \citep{Bolen:2004sq,Bambi:2007ty,Mignemi:2009ji,Chung:2017sdz}. 

In this paper, we investigate the possible effects of the GUP model
on the \textgreek{L}CDM cosmology. For this purpose, we use the entropic
force approach and the GUP model that used in \citep{Scardigli:2018jlm}. 

In Sec.2, we summarize the GUP-corrected gravity model based on \citep{Scardigli:2018jlm}.
Then in Sec.3, we give a brief description of the \textgreek{L}CDM
cosmology. In Sec.4, we find a Planck-scale-corrected Friedmann equations
based on the selected GUP model. We also note that this paper uses
the natural units where the speed of light $c$ and Boltzmann\textquoteright s
constant $k_{B}$ equal to one. The last section concludes the paper. 

\section{Quantum corrected entropic gravity}

One can derive the Unruh effect with the help of Heisenberg\textquoteright s
uncertainty principle when we assume that a photon has crossed the
Rindler event horizon \citep{Scardigli:1995qd,Gine:2018tjn}. If we
consider the Planck scale where the gravitational effect is neglected,
we know that there are some modifications of HUP named as generalized
uncertainty principle and extended uncertainty principle. So, these
modified models affect the structure of thermodynamical quantities
such as the Unruh effect. According to the paper \citep{Scardigli:2018jlm}
a GUP model is given as follows,
\begin{equation}
\Delta x\Delta p\geq\frac{\hbar}{2}\left[1+\beta\left(\frac{\Delta p}{m_{p}}\right)^{2}\right],\label{eq: GUP commutation}
\end{equation}
and

\begin{equation}
\left[\hat{x},\hat{p}\right]=i\hbar\left[1+\beta\left(\frac{\hat{p}}{m_{p}}\right)^{2}\right],\label{eq: GUP commutation-1}
\end{equation}
where the constant $\beta$ is used as a dimensionless deformation
parameter. By the help of this modified background, we can derive
the modified Unruh temperature as follows,
\begin{equation}
T\backsimeq T_{U}f\left(a\right).\label{eq: temp1}
\end{equation}
Here $T_{U}=ha/2\pi$ represents the ordinary Unruh temperature and
the function $f\left(a\right)$ is defined as

\begin{equation}
f\left(a\right)=\left[1+\frac{\beta}{2}\left(\frac{l_{p}a}{\pi}\right)^{2}\right].\label{eq: f(a)}
\end{equation}
Here, $a$ is the acceleration of the reference frame. According to
this background and using the Verlinde\textquoteright s entropic gravity
idea, a GUP-corrected gravitational field equation can be found as
follows \citep{Kibaroglu:2019odt},

\begin{equation}
R_{ab}-\frac{1}{2}g_{ab}R+\frac{\Lambda}{f\left(a\right)}g_{ab}=\frac{8\pi G}{f\left(a\right)}\left\{ T_{ab}+T_{ab}^{GUP}\right\} ,\label{eq: EFE_def}
\end{equation}
where $\Lambda$ represents the cosmological constant, $T_{ab}$ is
originated from ordinary matter and $T_{ab}^{GUP}$ comes from GUP
modification and its form, 

\begin{equation}
T_{ab}^{GUP}=\frac{1}{8\pi G}\left(\nabla_{a}\nabla_{b}f\left(a\right)-\frac{1}{2}g_{ab}\nabla^{2}f\left(a\right)\right).
\end{equation}
Moreover, in the concept of this GUP model, Newton\textquoteright s
second law of gravity takes the following form,

\begin{equation}
F=maf\left(a\right),
\end{equation}
where $f\left(a\right)$ is given by Eq.\ref{eq: f(a)}. 

\section{Review of \textgreek{L}CDM cosmology}

In this section, we present a brief review of the \textgreek{L}CDM
cosmological model which is also known as the standard model of cosmology
\citep{bergstrom2006cosmology}. According to this model, the Friedmann
equation is given the following form, 

\begin{equation}
H\left(t\right)^{2}=\frac{8\pi G}{3}\rho\left(t\right)-\frac{k}{a^{2}}+\frac{\Lambda}{3},
\end{equation}
and the acceleration equation is,

\begin{equation}
\frac{\ddot{a}}{a}=-\frac{4\pi G}{3}\left(\rho+3p\right)+\frac{\Lambda}{3},
\end{equation}
here $a\left(t\right)$ is a dimensionless arbitrary function of time,
known as the scale factor, which is related to the expansion of the
universe, $\rho\left(t\right)$ and $p\left(t\right)$ represents
total energy density and the pressure of cosmological fluids, $H\left(t\right)=\dot{a}\left(t\right)/a\left(t\right)$
is the Hubble parameter which describes the expansion rate of the
universe and the dot represents the time derivative of the corresponding
component. Moreover, the driving term $\Lambda/3$ is responsible
for the acceleration. Besides, the continuity equation can be given
as follows, 

\begin{equation}
\dot{\rho}+3H\left(\rho+p\right)=0
\end{equation}
Using these equations, the Raychaudhuri equation can be written as
follows,

\begin{equation}
\dot{H}=-4\pi G\left(\rho+p\right)+\frac{k}{a^{2}},
\end{equation}

\section{GUP corrected Friedmann equations}

We start with the Friedmann--Robertson--Walker universe with the
metric 

\begin{equation}
ds^{2}=dt^{2}-a\left(t\right)^{2}\left(dr^{2}+r^{2}d\Omega^{2}\right)
\end{equation}
where $a\left(t\right)$ is the scale factor, and $\Omega$ denotes
the line element of a unit sphere. According to Verlinde\textquoteright s
paper, we have a spherical holographic screen with a spatial region
which has the following physical radius,

\begin{equation}
\tilde{r}=a\left(t\right)r
\end{equation}
By assuming that our universe has homogeneity and isotropic form in
a large scale, the matter content of the universe forms a perfect
fluid with the following stress-energy tensor, 

\begin{equation}
T_{\mu\nu}=\left(\rho+p\right)u_{\mu}u_{\nu}+pg_{\mu\nu}\label{eq: T_mu_nu}
\end{equation}
and if we take the trace of this tensor we get,

\begin{equation}
T=T_{\mu}^{\mu}=\rho-3p\label{eq: T_trace}
\end{equation}
Here $u^{\mu}=\left(1,0,0,0\right)$ is the four-velocity and satisfies
the relation $g_{\mu\nu}u^{\mu}u^{\nu}=1$. Also $\rho\left(t\right)$
and $p\left(t\right)$ represents energy density and the pressure
of cosmological fluids, respectively. By using the conservation of
the energy--stress tensor, the continuity equation takes the following
form, 

\begin{equation}
\dot{\rho}+3H\left(\rho+p\right)=0
\end{equation}
The number of bits on the screen is defined as 

\begin{equation}
N=\frac{A}{G\hbar},
\end{equation}
where $A=4\pi\tilde{r}^{2}$ represents the area of the screen. According
to the equipartition law of energy, the total energy on the screen
can be written as follows, 

\begin{equation}
E=\frac{1}{2}Nk_{B}T.
\end{equation}
On the other hand, the energy can be represented as

\begin{equation}
E=M\label{eq: E=00003DM}
\end{equation}
where the mass $M$ represents the mass in the spatial region $V$.
Using Eq.\ref{eq: T_mu_nu} and Eq.\ref{eq: E=00003DM}, the total
mass in the spatial region can also be defined as 

\begin{eqnarray}
M & = & \int_{V}dV\left(T_{\mu\nu}u^{\mu}u^{\nu}\right)
\end{eqnarray}
where $T_{\mu\nu}u^{\mu}u^{\nu}$ is the energy density of the system.
Moreover, the acceleration of the radial observer at $r$ can be written, 

\begin{equation}
a_{r}=-\frac{d^{2}\tilde{r}}{dt^{2}}=-\ddot{a}r
\end{equation}
By using this result the modified Unruh temperature should be the
following form, 

\begin{equation}
T=\frac{\hbar a_{r}}{2\pi}f\left(a_{r}\right)=T_{U}f\left(a_{r}\right)
\end{equation}
From this background and by assuming the area as $A=4\pi\tilde{r}^{2}$and
the volume as $V=\frac{4}{3}\pi\tilde{r}^{3}$, we get 

\begin{equation}
\frac{\ddot{a}}{a}=-\frac{4\pi G}{3f}\rho,
\end{equation}
This equation represents the dynamical equations for the Newtonian
cosmology for the modified case. To obtain the Friedmann equations
for the GUP-deformed general relativity, we use active gravitational
mass (Tolman-Komar mass) $\mathcal{M}$ rather than total mass $M$
in the spatial region $V$ and it is defined as, 

\begin{eqnarray}
\mathcal{M} & = & 2\int_{V}dV\left(T_{\mu\nu}-\frac{1}{2}Tg_{\mu\nu}+\frac{\Lambda}{8\pi G}g_{\mu\nu}\right)u^{\mu}u^{\nu}
\end{eqnarray}
After taking this integral, we get the active gravitational mass as 

\begin{equation}
\mathcal{M}=\left(\rho+3p-\frac{\Lambda}{4\pi G}\right)V.
\end{equation}
Then using this new mass definition, we find the following equation,

\begin{equation}
\frac{\ddot{a}}{a}=-\frac{4\pi G}{3f}\left(\rho+p\right)+\frac{\Lambda}{3f}.\label{eq: F2_def}
\end{equation}
So, we find the acceleration equation including the cosmological term
for the dynamical evolution of the FRW universe. Multiplying a\.{ }a
on both sides of the last equation and using the continuity equation
in Eq.\ref{eq: T_mu_nu}, we get 

\begin{equation}
H^{2}=\frac{8\pi G}{3f}\rho\left(t\right)-\frac{k}{a^{2}}+\frac{\Lambda}{3f},\label{eq: H^2_def}
\end{equation}

Thus, we get the second Friedmann equation that is responsible for
the time evolution of the FRW universe. Here, k is the integration
constant and it can be interpreted as the spatial curvature of the
region V in Einstein\textquoteright s theory of gravity. Here, $k=1,0,-1$
correspond to a closed, flat, and open FRW universe, respectively.
The Eq.\ref{eq: H^2_def} can also be written as follows, 

\begin{equation}
H^{2}=\frac{8\pi G}{3f}\left(\rho-\rho_{\Lambda}\right)-\frac{k}{a^{2}},
\end{equation}
where an energy density corresponding to the cosmological constant
is defined as, 

\begin{equation}
\rho_{\Lambda}=-\frac{\Lambda}{8\pi G}
\end{equation}
In other words, it can be interpreted as an additional energy density
to the universe. Moreover, time derivation of the Hubble parameter
can be written as follows, 

\begin{equation}
\dot{H}=\frac{\ddot{a}}{a}-H^{2}
\end{equation}
Using this equation together with Eqs.\ref{eq: F2_def} and \ref{eq: H^2_def},
we get, 

\begin{equation}
\dot{H}=-\frac{4\pi G}{f}\left(\rho+p\right)+\frac{k}{a^{2}},
\end{equation}
Thus, we derive the GUP-corrected Raychaudhuri equation. With the
help of Eq.\ref{eq: H^2_def}, one can also write the matter density
as follows, 

\begin{equation}
\rho\left(t\right)=\rho_{c}+\rho_{k}+\rho_{\Lambda}
\end{equation}
where the critical density $\rho_{c}$ and the energy density corresponding
to the integration constant $\rho_{k}$ are defined as follows, 

\begin{equation}
\rho_{c}=\frac{3H^{2}}{8\pi G}f,\,\,\,\,\,\,\rho_{k}=\frac{3k}{8\pi Ga^{2}}f
\end{equation}
On the other hand, the density parameter is defined as $\Omega\left(t\right)=\rho/\rho_{c}$
and using this definition one can write,

\begin{equation}
\Omega\left(t\right)+\Omega_{k}\left(t\right)+\Omega_{\Lambda}\left(t\right)=1
\end{equation}
where,

\begin{equation}
\Omega_{k}\left(t\right)=-\frac{k}{H^{2}a^{2}},\,\,\,\,\,\,\Omega_{\Lambda}\left(t\right)=\frac{\Lambda}{3H^{2}f}
\end{equation}

These expressions represent density parameters with respect to the
integral constant and the cosmological constant, respectively. The
density parameter contains information about the shape of our universe.
If we suppose $\Omega\left(t\right)=1$, this model describes a flat
universe. The other conditions $\Omega\left(t\right)<1$ and $\Omega\left(t\right)>1$
corresponds to an open universe closed universe. Today it is known
that the value of the density parameter is close to one. From here,
we can write the following expression, 

\begin{equation}
\Lambda\simeq\frac{3k}{a^{2}}\left[1+\frac{\beta}{2}\left(\frac{l_{p}a}{\pi}\right)^{2}\right].\label{eq: cc_def}
\end{equation}

If we set the deformation parameter as $\beta=0$ the modified equations
go to their standard forms. Furthermore, some possible values of the
deformation parameter $\beta$, both gravitational and non-gravitational
cases, are summarized in \citep{Lambiase:2017adh}. According to this,
for gravitational cases, the values of $\beta$ change between $\beta<10^{21}$
and $\beta<10^{78}$. On the other hand, in some cases, $\beta$ could
get negative values \citep{Ong:2018zqn,Kanazawa:2019llj,Buoninfante:2019fwr,Buoninfante:2020guu}.
This idea may lead to interesting results because a negative deformation
parameter can change the sign of the cosmological constant in a special
condition. If we consider very high acceleration values and at appropriate
values, the sign of Eq.\ref{eq: cc_def} would be changed. 

\section{Conclusion}

In this paper, we aimed to find the possible contribution of the generalized
uncertainty principle on a cosmological model. We know that Verlinde\textquoteright s
entropic gravity idea provides a useful theoretical background to
extend/modify cosmological models. 

From this background, we used GUP-corrected gravitational model in
Eq.\ref{eq: EFE_def} (see also \citep{Kibaroglu:2019odt}) to derive
modified Friedmann equations. Eventually, we derived quantum corrected
Friedman equations Eq.\ref{eq: F2_def} and Eq.\ref{eq: H^2_def}.
Also, these equations can be seen as quantum corrected \textgreek{L}CDM
model. In a certain condition, if the deformation parameter $\beta$
in the function $f\left(a\right)$ in Eq.\ref{eq: f(a)} goes to zero,
the model reduces its conventional form. 

According to the function $f\left(a\right)$ and modified Friedmann
equations, the contribution that comes from GUP is very little because
the function contains a square of the Planck length. In this condition,
the corrections can be ignored when the Universe becomes large. But
there may be a considerable contribution for the early universe such
as the inflation phase. This is very important because the \textgreek{L}CDM
model is insufficient for the explanation the early Universe period. 

In addition, $f\left(a\right)$ could have a negative sign when we
chose negative deformation parameter and extreme acceleration conditions.
This may lead to change some signs of the resulting Friedmann equations
and the cosmological constant in Eq.\ref{eq: cc_def}. Thus, we can
say that this kind of deformation could affect the evolution of the
universe. 

Consequently, one can say that modifications of the entropic gravity
lead to derive extended cosmological models and these studies may
provide a deeper understanding of our Universe. 

\bibliographystyle{apsrev4-2}
\bibliography{gup_lambda_CDM}

\end{document}